\documentclass[fleqn,twoside]{article}
\usepackage[headings]{espcrc2}
\usepackage{epsf,epsfig}
\readRCS $Id: espcrc2.tex,v 1.2 2004/02/24 11:22:11 spepping Exp $
\def\slash#1{#1 \hskip-0.45em /}

\def\beq{\begin{eqnarray}}
\def\eeq{\end{eqnarray}}

\def\be{\begin{equation}}
\def\ee{\end{equation}}

\newcommand{\AmS}{{\protect\the\textfont2
  A\kern-.1667em\lower.5ex\hbox{M}\kern-.125emS}}

\title{Heavy-to-light form factors for non-relativistic
       bound states\thanks{
Talk presented at {\it QCD '05}, Montpellier, July 2005. \newline
       LMU-ASC 63/05, SI-HEP-2005-10}}

\author{G. Bell\address{
Arnold Sommerfeld Center, Department f\"ur Physik,
Ludwig-Maximilians-Universit\"at M\"unchen,
D-80333 M\"unchen, Germany}%
       \ and 
        Th.~Feldmann\address{
Fachbereich Physik,
Universit\"at Siegen,
D-57068 Siegen, Germany}%
}
       
\begin{document}

\begin{abstract}
We investigate transition form factors between non-relativistic QCD bound
states at large recoil energy. Assuming the decaying quark to be much
heavier than its decay product, the relativistic dynamics can be
treated according to the factorization formula for heavy-to-light 
form factors obtained from the heavy-quark expansion in QCD.
The non-relativistic expansion 
determines the bound-state wave functions to be Coulomb-like.
As a consequence, one can explicitly calculate the so-called 
``soft-overlap'' contribution to the transition form factor.
\end{abstract}

\maketitle


\section{Introduction}

The factorization of high-energy hadronic processes into 
perturbative QCD scattering amplitudes and universal hadronic
parameters is one of the most powerful tools in strong-interaction
physics. In the case of massive $B$\/-meson decays into light
energetic hadrons, this factorization is complicated due to the
presence of three relevant scales: (i) a ``\underline{h}ard'' scale set by
the mass $m_b$ of the decaying $b$\/-quark, (ii) a ``\underline{s}oft'' (``\underline{c}ollinear'') 
scale $\Lambda \ll m_b$ set by the non-perturbative dynamics in the initial (final) hadronic bound states, and (iii) a ``\underline{h}ard-\underline{c}ollinear'' scale $\mu_{\rm hc} \sim \sqrt{m_b \Lambda}$ with $\Lambda \ll \mu_{\rm hc} \ll m_b$ 
which appears due to interactions between the low-energetic particles
in the initial and the high-energetic particles in the final state.

A systematic separation of these scales in exclusive two-body
$B$\/-decays has been worked out in \cite{BBNS}. The quantum-field theoretical
formalism has been developed in terms of a so-called ``soft-collinear
effective theory'' (SCET) in \cite{SCETa,Beneke:2002ph}.
In the following we consider the form factor for a $B$\/-meson transition into a single energetic light hadron, for instance a pion. Schematically, the factor\-iza\-tion formula in the heavy-quark limit, $m_b \to \infty$, reads~\cite{Beneke:2000wa}
\be
\langle \pi| \bar \psi \, \Gamma_i \, b |B\rangle
\simeq  T_i^{\rm I} \cdot \xi_\pi +
   T_i^{\rm II}  \otimes \phi_B
   \otimes \phi_\pi
\label{factorization}
\ee
Here, the short-distance function $T_i^{\rm I}$ contains dynamics from the
hard scale, whereas $T_i^{\rm II}$ contains hard and hard-collinear dynamics.
The second term in (\ref{factorization}) is thus completely factorized, and
the light-cone distribution amplitudes $\phi_B$ and $\phi_\pi$, which provide the
non-perturbative information, contain dynamics below $\mu_{\rm hc}$, only. This
is not the case for the first term in (\ref{factorization}) where the so-called
``soft'' form factor $\xi_\pi$ still contains dynamics at the hard-collinear
scale and cannot be factorized further. 
These non-factorizable contributions are related to endpoint divergences
which appear when one of the partons in the perturbative scattering amplitude
has almost vanishing (light-cone) energy, see the discussion in
\cite{Bauer:2002aj,Beneke:2003pa,Lange:2003pk}. In (\ref{factorization}) the 
non-factorizable form factor $\xi_\pi$ does not depend on the Dirac
structure $\Gamma_i$ of the $b$\/-quark decay. As a consequence,
heavy-to-light form factors obey symmetry
relations \cite{Charles:1998dr} which 
are broken by calculable perturbative effects in $T_i^{\rm I,II}$
in the heavy-quark limit \cite{Beneke:2000wa,Becher:2004kk,Beneke:2004rc}.

The idea of the presented work (see also \cite{guido}) 
is to extend this formalism to non-relativistic bound states where, in the limit of small relative velocities, the non-perturbative dynamics is contained in the usual Coulomb wave function. Consider, for instance, the decay 
$B_c \to \eta_c \ell\nu$ where we assume the hierarchy of scales
\be
  \Lambda_{\rm QCD} 
\ll m_c \ll \sqrt{m_c m_b} \ll m_b
\ee
The above formalism can be applied if one identifies the
charm-quark mass $m_c$ with the soft scale. The latter is larger than the QCD scale, and the dynamics at $m_c$ is still perturbative,
such that the soft (non-factorizable) form factor can be calculated 
explicitly. The non-relativistic approach may also be considered
as a toy model for the $B \to \pi$ case which may shed further light 
on the structure of the factorization formula~(\ref{factorization}).


\section{Non-relativistic bound states}

\begin{figure}[tbhp]
 \centerline{
  \epsfclipoff
  \epsfig{file=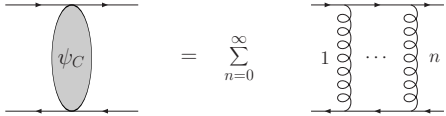, width=0.37\textwidth,bb=100 640 435 705}
 }
\caption{Resummation of "potential" gluons into a 
non-relativistic Coulomb wave function.}
\label{fig:coulomb}
\end{figure}

The wave function for a non-relativistic bound
state can be obtained from the resummation of 
"potential" gluon exchange, see Fig.~\ref{fig:coulomb}. 
The solution of the corresponding Schr\"odinger equation with Coulomb potential
describes a wave function which is sharply peaked around 
small three-momenta $\vec p = {\cal O}(m_r \alpha_s )$
where $m_r$ is the reduced quark mass.
The normalization of the wave function gives the meson decay constant. 

In this picture, the $B_c$ meson is described
as a bound state of a heavy bottom quark with mass $M=m_b$ and a lighter 
charm quark with mass $m=m_c$. Consequently,
to first approximation with respect to a simultaneous
expansion in $m/M$ and $v = |\vec p|/m_r$,
the $B_c$ meson consists of a heavy quark with momentum $M w_\mu$ 
and a light quark with momentum $m w_\mu$, where $w^\mu$ is the four-velocity
of the heavy meson ($p_\mu \simeq M w_\mu$). 
The spinor degrees of freedom for the heavy meson
in the initial state are represented by the Dirac projection
$
  \frac{1 + \slash w}{2} \, \gamma_5.
$
Similarly, the $\eta_c$ meson is represented as a
$c\bar c$ bound state where both constituents have
approximately equal momenta $m w'_\mu$, and the
total momentum of the meson is $p'_\mu = 2 m w'_\mu$.
The Dirac projection for the $\eta_c$ meson in the final state
is thus given by
$  
  \frac{1 - \slash w'}{2} \, \gamma_5 \ .
$


\section{Relativistic dynamics}

In the following, for simplicity, we will consider the case of maximum
recoil energy, i.e.\ the momentum transfer is given by
$
  q^2 = (p - p')^2  =  0.
$
Furthermore, we will concentrate on the form factor
$F_+$, defined via
\be
  \langle \eta_c|\bar c \, \gamma_\mu \, b|B_c\rangle
= F_+(q^2) \, (p_\mu + p'_\mu) + F_-(q^2) \, q_\mu
\ee
According to the general discussion,
we have to consider hard, hard-collinear, collinear and 
soft gluon exchange in order to describe the relativistic dynamics of the 
transition form factor.
Because of the large energy transfer, there is no direct overlap between 
the non-relativistic wave functions in the heavy-quark limit.

\subsection{Spectator scattering at tree level}

\begin{figure}[tbhp]
 \centerline{
  \epsfclipoff
  \epsfig{file=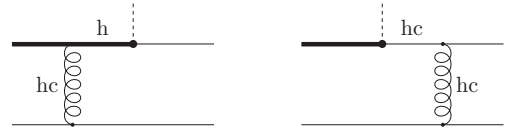, width=0.42\textwidth, bb=  175 650 425 715}
 }
\caption{Tree-level exchange of one hard-collinear gluon. The
 thick line represents the heavy quark.}
\label{fig:tree}
\end{figure}

At tree level we have to calculate the two diagrams
in Fig.~\ref{fig:tree} where a hard-collinear gluon
is exchanged between the active quarks and the spectator. 
The result for the form factor reads
\be
F_+^{\rm LO}(0)
=
\frac{4 \pi \alpha_s(\mu) C_F}{N_c}
\,  \frac{3 f_m f_M}{m M} 
\ee
where $f_M$ and $f_m$ denote the (non-relativistic) decay constants
of the initial and final-state meson, respectively.
Notice, that to this (fixed) order in perturbation theory,
one does not encounter endpoint singularities, because, in
contrast to the relativistic case, the light-cone wave functions
of the non-relativistic bound states have vanishing support at
the endpoints. As we will see below, the endpoint configurations
require the exchange of at least another relativistic 
gluon (soft or collinear, respectively).

\subsection{Spectator scattering at one-loop}

\begin{figure}[tbhp]
   \centerline{
  \epsfig{file=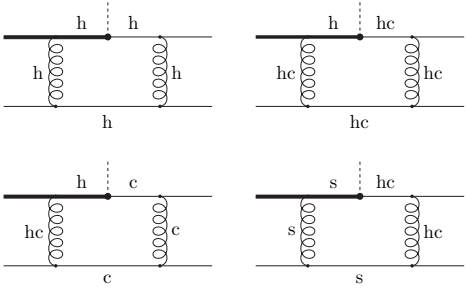, width=0.39\textwidth, bb=165 570 435 715}
 }
\caption{Hard, hard-collinear, collinear and soft momentum regions for the pentagon diagram.}
\label{fig:pentagonregions}
\end{figure}

At order $\alpha_s^2$ the form factor receives
contributions from various 1-loop diagrams (details will
be given in \cite{wip}).
Let us consider, as an example, the pentagon diagram shown in
Fig.~\ref{fig:pentagonregions}. It is instructive to separate 
the various momentum regions \cite{Smirnov:1998vk} that
contribute when the diagram
is calculated within dimensional regularization ($D = 4 -2\epsilon$). 
In our example, the hard region is power-suppressed.
The remaining regions contribute to the NLO corrections
\be
 \Delta F_+^{\rm NLO} \big|_{\rm pent.} =  F_+^{\rm LO} 
   \, \frac{\alpha_s C_F}{4\pi} \left\{ I_{\rm hc} + I_{\rm s} + I_{\rm c}
\right\} \ .
\ee
In Feynman gauge one has
\beq
  I_{\rm hc} &=&
   \frac13 \left( \frac{1}{\epsilon^2} + \frac{1}{\epsilon}
      \ln \frac{2\mu^2}{m M} + \frac12 \ln^2 \frac{2 \mu^2}{m M} \right)
\cr && 
    - \frac{1}{\epsilon} - \ln \frac{2 \mu^2}{m M}
    - \frac23 \ln 2 - \frac{\pi^2}{36} + \frac43
\label{eq:hc}
\ , \nonumber \\[0.2em]
  I_{\rm c} &=& 
    \frac43 \left( \frac{1}{\epsilon} + \ln \frac{\mu^2}{m^2} \right)
    + \frac{11}{3} \ln 2 - 2 + \frac{\pi^2}{9} 
\cr && 
    - \frac13 \left( \frac{1}{\delta} + \ln \frac{2\nu^2}{M m}\right)
      \left( \frac{1}{\epsilon} + \ln \frac{\mu^2}{m^2} + 1 \right)
\ , \nonumber \\[0.2em]
  I_{\rm s} &=& 
    -  \frac13 \left( 
      \frac{1}{\epsilon^2} + \frac{1}{\epsilon}
      \ln \frac{\mu^2}{m^2} + \frac12 \ln^2 \frac{\mu^2}{m^2} \right)
   \cr && 
    + \frac53 \left(\frac{1}{\epsilon} + \ln \frac{\mu^2}{m^2}\right)
    + \frac{5\pi^2}{36} - \frac83 
  \cr && 
    + \frac13 \left( \frac{1}{\delta} + \ln \frac{\nu^2}{m^2}\right)
      \left( \frac{1}{\epsilon} + \ln \frac{\mu^2}{m^2} + 1 \right)
\eeq
In the soft and collinear momentum region we had to 
introduce an additional regularization procedure for
the divergences related to light-cone momentum
fractions of the internal loop momenta approaching the
endpoints. We apply an analytic continuation
of one of the gluon propagators, 
$1/[k^2+i\eta] \to (-\nu^2)^\delta /[k^2+i\eta]^{1+\delta}$. In 
this way endpoint divergences show up as poles in $1/\delta$.

In fixed-order perturbation theory, one would simply
add up all three momentum regions at a common renormalization scale
$m \leq \mu \leq M$. 
This reproduces the divergences of the full QCD result, whereas the
endpoint divergences and the dependence on the ad-hoc parameters
$\nu^2$ and $\delta$ vanish.
It is instructive to separate the contributions related
to the endpoint divergences in the sum of 
soft and collinear momentum regions
\beq
  I_{s+c}\big|_{\rm endpoint} &=&
 \frac13  \ln \frac{M}{2m}
      \left( \frac{1}{\epsilon} + \ln \frac{\mu^2}{m^2} + 1 \right)
\label{endpoint}
\eeq
The point to notice is that the result contains 
large logarithms, $\ln M/m $,
which involve the ratio of the two distinct physical scales.
In renormalization-group-improved perturbation theory one would like 
to resum these large logarithms into short-distance 
coefficient functions. Whether this is possible can be read off the 
individual momentum regions:

The hard-collinear integral $I_{\rm hc}$ in (\ref{eq:hc}) does not contain 
large logarithms if one chooses a renormalization scale 
$\mu \sim \sqrt{m M}$.
For the soft and collinear integrals $I_{\rm s,c}$
the choice $\mu \sim m$, which refers to the typical virtualities of soft
and collinear fields, leads to small logarithms $\ln \mu^2/m^2$.
Still, one would be left with the large logarithm  $\ln M/m$ in 
(\ref{endpoint}). This implies that, in general,
one {\em cannot}\/ resum all large logarithms $\ln m/M$
into short-distance (hard-collinear) coefficient functions. 
In the context of the effective-theory approach such contributions 
are thus identified as non-factorizable \cite{Beneke:2003pa,Lange:2003pk}. 
 
In the $B \to \pi$ case, the light quark mass enters as an IR regulator
of the order of $\Lambda_{\rm QCD}$, and the sum of soft and collinear
momentum regions define a {\em non-perturbative}\/ form factor which 
still depends on the heavy-quark mass in an unknown, non-analytic way.
In contrast, for the non-relativistic description of the $B_c \to \eta_c$ 
transition, the light charm quark mass is a physical parameter,
and the soft and collinear momentum regions still refer to the
perturbative QCD sector.


\section{Numerical applications}

For the numerical estimate we stick to fixed-order perturbation
theory, summing up all diagrams at order $\alpha_s$.
We do not attempt to resum large logarithms
(a more thorough discussion will be given in \cite{wip}).
Without resummation our result shows a sizeable but moderate 
dependence on the renormalization scale $\mu$. 
We emphasize that this dependence would become 
out of control in the $B \to \pi$ case where
the large logarithms have to be 
counted as $1/\alpha_s$.
\begin{figure}[hbtbhp]
\centerline{\epsfclipoff
\epsfig{file=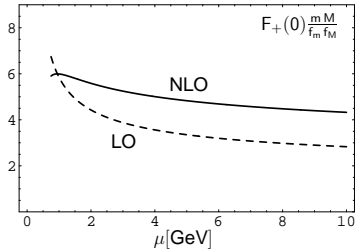, width=0.29\textwidth, bb=155 575 375 690}
}
\caption{Renormalization-scale dependence 
of $F_+(0)$ for $B_c \to \eta_c$ transitions at LO and NLO.}
\label{fig:mudep}
\end{figure}

In Fig.~\ref{fig:mudep}
we compare the renormalization-scale dependence of the LO and NLO result, using
$M=5$~GeV, $m=1.5$~GeV, $\alpha_s(M)=0.2$, 
$\ln\left[\alpha_s(m)/\alpha_s(M)\right] \simeq 0.4$
as an example. We observe that the
NLO corrections stabilize the perturbative expansion, 
with the scale-uncertainty
improving from about 30\% to 15\%.

It is possible to disentangle the non-factorizable and factorizable
form factor contributions, using the definitions in \cite{Beneke:2002ph}. 
It turns out that, by accident (i.e.\ not due to an obvious symmetry),
 the non-factorizable part of the form factor $F_+(0)$ 
{\em numerically}\/ dominates over the factorizable one by
about a factor of 5.  
Notice that in the non-relativistic set-up both,
the factorizable and non-factorizable contributions, are of order $\alpha_s(\mu)$.


\section{Summary}

We have shown how to describe the dynamics of non-relativistic bound states 
in heavy-to-light quark decays at large recoil, using the example of
$B_c \to \eta_c \ell \nu$, where we consider $m_c \ll m_b$. We have worked 
to leading order in the non-relativistic and the $m_c/m_b$ expansion, and took
into account NLO effects from relativistic gluon exchange. We found that
(without resummation of large logarithms) the perturbative uncertainties of
the NLO result amount to about 15\%. This is already below the expected size
of power-corrections (about 30\%)
which should be included for reliable phenomenological estimates.

\subsection*{Acknowledgements}

We thank Martin Beneke for many helpful discussions.

\end{document}